\documentclass[%
reprint,
 amsmath,amssymb,
 aps,
 prfluids,
]{revtex4-2}

\usepackage{graphicx}
\usepackage{dcolumn}
\usepackage{bm}

\usepackage{mathtools}

\usepackage{hyperref}
\usepackage{xcolor}
\usepackage[caption=false]{subfig}

\usepackage{cleveref}
\usepackage{rotating}
\usepackage{adjustbox}
\usepackage{tikz}
\usetikzlibrary{shapes}

\usepackage{etoolbox}
\usetikzlibrary{plotmarks}
\usepackage{MnSymbol}

\usepackage{comment}
\usepackage{epstopdf}

\newcommand\norm[1]{\left\lVert#1\right\rVert_2}
\newcommand*\diff{\mathop{}\!\mathrm{d}}
\newcommand{\avg}[1]{\left<#1\right>}

\DeclareRobustCommand\full  {\tikz[baseline=-0.6ex]\draw[thick] (0,0)--(0.5,0);}
\DeclareRobustCommand\dotted{\tikz[baseline=-0.6ex]\draw[thick,dotted] (0,0)--(0.54,0);}
\DeclareRobustCommand\dashed{\tikz[baseline=-0.6ex]\draw[thick,dashed] (0,0)--(0.54,0);}

\DeclareRobustCommand\dashdot {\tikz[baseline=-0.6ex]\draw[thick,dash dot] (0,0)--(0.5,0);}

\definecolor{darkgreen}{RGB}{0.0, 100, 0.0}
\definecolor{purple}{RGB}{128, 0.0, 128}
\definecolor{orange}{RGB}{255, 149, 0.0}
\definecolor{grey}{RGB}{127, 127, 127}

\definecolor{blueplot}{RGB}{12.0, 93, 165}
\definecolor{greenplot}{RGB}{0.0, 185, 69}
\definecolor{orangeplot}{RGB}{255, 149, 0.0}
\definecolor{brownplot}{RGB}{140, 86, 75}
\definecolor{pinkplot}{RGB}{227, 119, 194}

\newcommand\marksymbol[3]{\tikz[draw = #2, fill = white, scale = #3]\pgfuseplotmark{#1};}
\newcommand\marksymbolfull[3]{\tikz[draw = #2, fill = #2, scale = #3]\pgfuseplotmark{#1};}
\newcommand\diams[2]{\tikz[draw = #1, scale = #2]\pgfuseplotmark{diamond};}
\newcommand\diamsfull[2]{\tikz[draw = #1, fill = #1, scale = #2]\pgfuseplotmark{diamond*};}

\newcommand\DownTriangle{
\begin{tikzpicture}
 \node[regular polygon,regular polygon sides=3, draw, shape border rotate=180]{};
\end{tikzpicture}
}
\newcommand\RightTriangle{
\begin{tikzpicture}
 \node[regular polygon,regular polygon sides=3, draw, fill = pinkplot, shape border rotate=270]{};
\end{tikzpicture}
}

\begin{document}

\title{Behind the mirror: the hidden dissipative singular solutions of ideal reversible fluids on log-lattices}
\author{Guillaume Costa}
\author{Amaury Barral}
\author{Adrien Lopez}
\author{Quentin Pikeroen}
\author{Berengere Dubrulle}
 \email{berengere.dubrulle@cea.fr}
\affiliation{%
 Université Paris-Saclay, CEA, CNRS, SPEC, 91191, Gif-sur-Yvette, France
}%

\date{\today}

\begin{abstract}
Empirical observations show that turbulence exhibits a broad range of scaling exponents, characterizing how the velocity gradients diverge in the inviscid limit. These exponents are thought to be linked to singular solutions of the Euler equations. In this work, we propose a dynamic approach to construct concept of these solutions directly from the fluid equations, using a reversible framework and introducing the efficiency $\cal{E}$, a non-dimensional number that quantifies the amount of energy stored within the flow due to an applied force. To circumvent the computational burden of tracking singularities at finer and finer scale, we test this approach on fluids on log-lattices, which allow for high effective resolutions at a moderate cost, while preserving the same symmetries and global conservation laws as ordinary fluids. We observe a phase transition at a given efficiency, separating regular, viscous solutions (hydrodynamic phase), from singular, inviscid solutions (singular phase). The singular solutions experience self-similar blow-ups with exponents corresponding to non-dissipative solutions. By applying a stochastic regularization, we are able to go past the blow-up, and show that the resulting solutions converge to power-law solutions with exponents characterizing dissipative solutions. Overall, the range of scaling exponents observed for log-lattice solutions is comparable to those of ordinary fluids.
\end{abstract}

\maketitle
\noindent\makebox[0pt][r]{%
  \begin{tikzpicture}[remember picture, overlay]
    \node[anchor=north east, xshift=-1cm, yshift=-0.5cm] at (current page.north east) {%
      \footnotesize\textbf{Submitted to \textit{Phys. Rev. Fluids}}};
  \end{tikzpicture}%
}


\section{\label{sec:Intro}Introduction}
 In 1949, Onsager postulated the existence of solutions to the incompressible Navier-Stokes equations (NSE) that can dissipate energy even in absence of viscosity, provided they are sufficiently irregular i.e; with a spectral exponent less than the Kolmogorov value~\cite{onsager1949statistical}. 
  This conjecture was proved by~\cite{Buckmaster18}, who used convex integration to construct arbitrary many weak dissipative solutions to the Euler equations (i.e. {\sl inviscid} NSE). However, the relevance of these constructions to ordinary fluid remains unclear. On the one hand, multi-fractal analysis of experimental~\cite{A96} and numerical~\cite{faller2022dissipation} data reveals a wide range of possible multi-fractal exponents characterizing how velocity gradients diverge in the inviscid limit.
  Applying Isett's theorem~\cite{Isett18}, one can show that these exponents correspond to both dissipative and non-dissipative singular solutions of the Euler equations. On the other hand, the lack of a dynamical construction of these solutions (i.e., derived directly from the fluid equations) prevents the identification of the mechanisms responsible for the selection of a given solution. In particular, it is still not clear how to connect dissipative or non-dissipative singular solutions of the Euler equations with solutions of the NSE at arbitrarily large Reynolds numbers. In addition to these mathematical difficulties, there are significant computational hurdles: exploring the inviscid limit of the NSE requires extremely high resolutions that are beyond the capabilities of current supercomputers. One may then consider simpler systems, that conserve all the symmetries and global conservation laws of the NSE and the Euler equations, while retaining a low computational cost. In this article, we consider the simpler model of fluid on log-lattices (LL), introduced by~\cite{campolina2019fluid, campolina2021fluid}, which is based on a controlled decimation of Fourier modes in d-dimensions. Their dynamical equation is obtained by projecting the fluid equations onto a sparse Fourier grid, consisting of exponentially spaced nodes. Previous simulations of LL-Euler equations have evidenced the existence of a self-similar blow-up, during which the energy spectrum scales like $E(k)\sim k^{-7/3}$~\cite{campolina2018chaotic, pikeroenSingularite}. On the other hand, simulations of forced LL-NSE at very low viscosity produces stationary solutions with a Kolmogorov energy spectrum $E(k)\sim k^{-5/3}$~\cite{campolina2021fluid,pikeroenSingularite}. This difference is not surprising from a physical standpoint, as the two equations do not share the same symmetries. In particular, the Euler equations are time-reversal symmetric, whereas the NSE are not. In this article, we show that by restoring the time-reversal symmetry, we can build a continuous family of solutions indexed by one parameter, the efficiency ${\cal E}$, that continuously bridges LL-Euler solutions with viscous LL-Navier-Stokes solutions. This family of solutions encompasses both dissipative and non-dissipative singular solutions to the Euler equations, thereby revealing otherwise hidden solutions.

\section{\label{sec:Framework} Reversible Navier-Stokes equations and Log-lattices projections}
    \subsection{\label{sec:RNS}Reversible Navier-Stokes equations}
        Restoring the time reversal symmetry in viscous fluids can be achieved in several ways, one of which originates from Gallavotti~\cite{gallavotti1996equivalence}. The idea consists in introducing a time-dependent viscosity $\nu_r[\bm{u}(t)]$, odd in $\bm{u}$, such that the NSE is left invariant under the transformation $\mathcal{T}: t\rightarrow-t; \bm{u}\rightarrow-\bm{u}; p\rightarrow p$. Specifically, one can define the reversible viscosity $\nu_r$ to ensure conservation of the total kinetic energy $E$~\cite{shukla2019phase, costa2023reversible} at all time. The resulting equation is called reversible Navier-Stokes equation (RNSE) and reads:
        \begin{equation}
            \begin{aligned}
                \nabla \cdot \bm{u} &= 0, \\
                \partial_t \bm{u} + (\bm{u} \cdot \bm{\nabla}) \bm{u} &= -\bm{\nabla} p + \nu_r \bm{\Delta} \bm{u} + \bm{f}, \\
                \nu_r[\bm{u}] &= \frac{\int_{\mathcal{D}} \bm{f} \cdot \bm{u} \, \diff \bm{x}}{\int_{\mathcal{D}} \norm{\bm{\nabla} \times \bm{u}}^2 \, \diff \bm{x}}
            \end{aligned}
            \label{eq:RNS}
        \end{equation}
        where $\bm{u}$ is the velocity, $p$ is the pressure, $\bm{f}$ is a force. We have set the constant density equal to $1$. The symmetry of this equation under the previously mentioned time reversal $\mathcal{T}$ is indeed ensured by the odd nature of $\nu_r$ with respect to $\bm{u}$. Despite the fluctuating nature of the viscosity, fluids obeying the RNSE adhere to Richardson's phenomenology, in which work is converted into heat through the creation of progressively finer energy-transporting structures (eddies) up to their breaking at the Kolmogorov length $\eta$~\cite{shukla2019phase}. In the process, a finite amount of mechanical energy is stored within the fluid, through erratic movements of eddies of all sizes. The efficiency of this storage can be quantified by the non-dimensional number ${\cal E}$~\cite{Lopez2025}. Defining spatial average of any function $\psi(x,t)$ as $\langle\psi \rangle=(\int d\bm{x})^{-1}\int d\bm{x} \psi$, we can express the efficiency $\cal{E}$ as:
        \begin{equation}
            {\cal E}=\frac{E}{L_f f_0},
            \label{eq:useful}
        \end{equation}
        where $L_f$ is a characteristic scale of forcing, defined e.g. as $(2\pi/L_f)^2=(1/3)\langle \nabla \bm{f}\cdot \nabla \bm{f}\rangle/\langle \bm{f}\cdot \bm{f}\rangle$, $E=\langle \bm{u}\cdot \bm{u}\rangle$ and $f_0=\langle \bm{f}\cdot \bm{f}\rangle^{1/2}$. In the RNSE framework, $E$ and $f_0$ are external input parameters, so that the efficiency ${\cal E}$ can be used as a control parameter. By convention, Euler solutions are such that $\nu=f_0=0$, corresponding to an infinite efficiency ${\cal E}=\infty$.
        
    \subsection{\label{sec:LL} Log-lattices}
\begin{figure}[!htb]
        	\centering
        	\includegraphics[width=0.9\columnwidth]{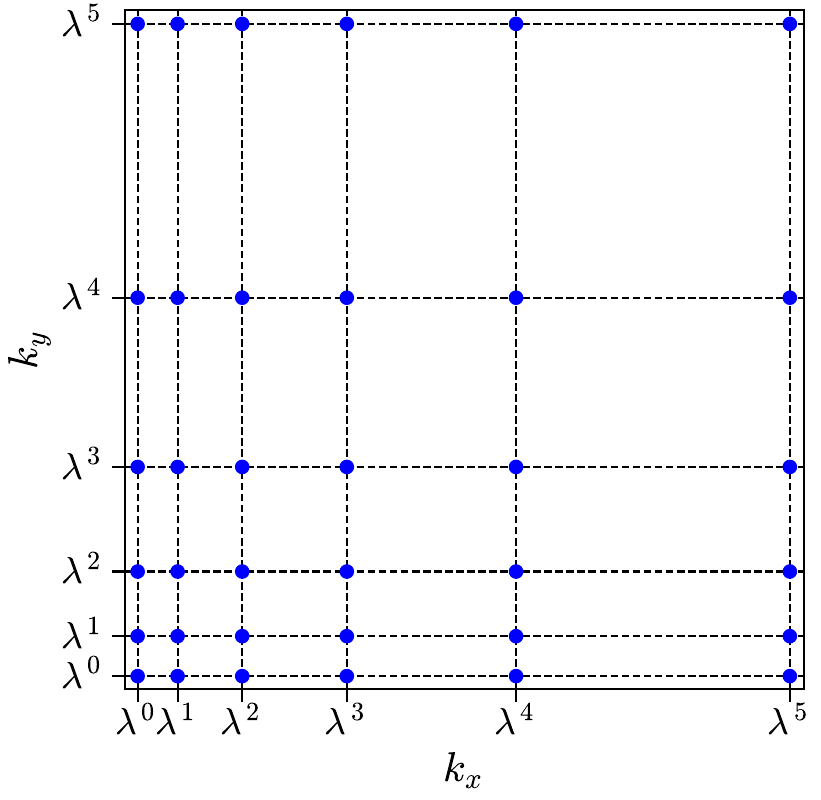}
        	\caption{Example of 2D log-lattice with spacing parameter $\lambda$.}
        	\label{fig:Grid}
        \end{figure}

        Log-lattices (LL) are discretized logarithmic grids, where the modes evolve according to a power-law given by the equation:
        \begin{equation*}
            k_n= k_0\lambda^n,
        \end{equation*}
        where $\lambda$ is the log-lattice spacing parameter. A 2D example of a log-lattice is given in Figure~\ref{fig:Grid}. Computations on such lattices are carried out by projecting the equation of interest onto the lattice in Fourier space.
        This construction is described in more detail in Campolina and Mailybaev~\cite{campolina2021fluid, campolina2019fluid}.
    
        \medbreak\noindent As all calculations on LL are purely spectral, the non-linear term is associated with a convolution product which couples modes in triadic interactions such that:
        \begin{equation}
            \label{eq:triadic}
            \lambda^m=\lambda^n + \lambda^q \text{, (m,n,q)} \in \mathbb{Z}^3.
        \end{equation} 
        This equation admits solutions, which restrict the possible values of $\lambda$ to three families, each having $z$ interactions in $D$ dimensions:
    
        \begin{itemize}
            \item $\lambda = 2$ ($z=3^D$).
            \item $\lambda = \sigma \approx 1.325$, the plastic number ($z=12^D$)
            \item $\lambda$ such that $1 = \lambda^b -\lambda^a$ for some integers $0< a < b$.
            ($a,b) \ne (1,3), (4,5)$ with $\gcd(a,b)=1$ ($z=6^D$).
        \end{itemize}
        The lowest possible values of $a$ and $b$, which are ($1,2)$, imply that the spacing parameter $\lambda$ corresponds to the golden ratio ($\phi \approx 1.618$).
        
        \medbreak\noindent LL are endowed with a comprehensive mathematical framework that supports a wide range of computations. For more information, refer to Campolina and Mailybaev~\cite{campolina2021fluid, campolina2019fluid}.
    
    \subsection{\label{sec:Num}Numerical details}
        All presented simulations are performed using a scaling parameter $\lambda = \phi$. 
        The integration is carried out in three steps. Starting from the initial conditions $\bm{\hat{u}}(t)$, we first solve the equation in its inviscid form using an explicit adaptive Runge-Kutta method of order 4--5. This gives us $\bm{\hat{u}}(t+dt)_{\nu=0}$ where $dt$ is the time-step. For RNSE, we then compute the reversible viscosity $\nu_r$ such that the total kinetic energy $E$ is conserved.
        
        \medbreak\noindent Finally, we apply the chosen viscosity using a technique similar to viscous splitting: $\bm{\hat{u}}(t+dt)=\bm{\hat{u}}(t+dt)_{\nu=0}e^{-\nu k^2 dt}$.
    
        \noindent Note that the forcing term is chosen as follows:
        \begin{equation}
            \label{eq:forc}
            \begin{split}
                \hat{f}_x(\textbf{k}) &= f_0 \text{ if } 15 < \norm{\bm{k}} <16 \text{ else } 0 , \\
                \hat{f}_y(\textbf{k}) &= f_0 \text{ if } 15 < \norm{\bm{k}} <16 \text{ else } 0 ,\\
                \hat{f}_z(\textbf{k}) &= 0.
            \end{split}
        \end{equation}
        Note that the form of the forcing has little impact on the results due to the very general definition of the control parameter e.g. the efficiency (Eq.~\ref{eq:useful})
        For blow-up simulations, we use an \textit{adaptative algorithm} to circumvent the truncation effect. Specifically, we add an additional mode to the grid whenever the energy in the last shell exceeds a certain threshold. In our simulations, this threshold was set to $E_{\text{threshold}} = 1e-200$.

    \subsection{Stochastic regularization protocol}
        \subsubsection{Setup}
            We first perform a blow-up simulation using an adaptative grid (see \textit{Numerical details}) with the desired parameters. Time steps at various resolutions $N$ are saved and stored. These are then used as initial conditions to launch regularized simulations at a fixed resolution $N_* = N_b(t)$ corresponding to the instantaneous resolution of the blow-up simulation at time $t$.
    
        \subsubsection{Protocol description}
            In order to pass the blow-up time $t_b$, one needs to \textit{stabilize} the solution. This can be achieved using different protocols, with the most widely used being (i) fixing the resolution to $N_*$ and letting the system evolve (ii) adding a small viscosity $\nu$.
            
            \medbreak\noindent While (i) preserves all the properties of the original equation, it is well known that thermalization effects arise from Galerkin truncation~\cite{krstulovic2008two}. These effects first lead to an accumulation of energy at the smallest accessible scales, effectively altering the system's dynamics locally (at small scales, $E(k) \propto k^{-\alpha}, \alpha < 5/3$),  which then propagates backward, contaminating the initial range.  It is therefore unclear whether the obtained results are a pure product of the Galerkin truncation or are physically relevant, protocol (i) was consequently discarded in this article.
    
            \medbreak\noindent The second protocol (ii) presents two issues. First, it introduces a change in the equation (e.g., switching from the Euler equations to the NSE), which results in a change in the dynamics at all scales. The second issue is not universal to every system but is specific to the nature of the RNSE. Indeed, since the viscosity is no longer a constant in the RNS system but an evolving quantity, one cannot simply add a viscosity, no matter how small, into the system, as this would break the RNS energy conservation scheme.
            
            \medbreak\noindent For these reasons, we decided to implement a different kind of protocol based on the addition of a multiplicative noise to pass the blow-up time $t_b$ without thermalizing.  
    
            \medbreak\noindent The protocol consists of two steps: (i) perform a time-step, using the unchanged equation of the considered system, and (ii) apply the transformation $\Gamma$ to the obtained velocity field, where $\Gamma$ is defined as follows:
            \begin{equation*}
                 \begin{split}
                    & \Gamma u(\bm{k}, t) = u(\bm{k}, t)\eta(\bm{k},t), \\
                    & \eta(\bm{k},t) = 
                        \left\{\begin{matrix}  X_{\bm{k}}(t), \text{ if } \bm{k} \in S_{N_*},
                                 \\  1, \text{ else}. 
                        \end{matrix}\right.
                 \end{split}
             \end{equation*}
             where $S_{N_*}$ denotes the last shell i.e modes such that $k_0 \lambda^{N_* - 2} \leq \norm{\bm{k}} < k_0\lambda^{N_* - 1}$ and $\forall t\; X_{\bm{k}}(t) \sim \mathcal{U}[-1, 1]$.
    
            \medbreak\noindent The solved equation remains unchanged $\forall \bm{k} \in S_i, \; i < N_*$ and is only slightly perturbed in $S_{N_*}$. Additionally, the regularization vanishes as $N_* \rightarrow \infty$ as the energy in the last mode approaches 0 in the infinite grid limit.
    
            \medbreak\noindent To some extent, our protocol shares similarties with the 1D tygers purging method proposed by~\cite{murugan2020suppressing} although it differs by being a continuous, 3D protocol.
    
        \subsubsection{Convergency}
            \begin{figure} [!htb]   
                            {\includegraphics[width =.9\columnwidth]{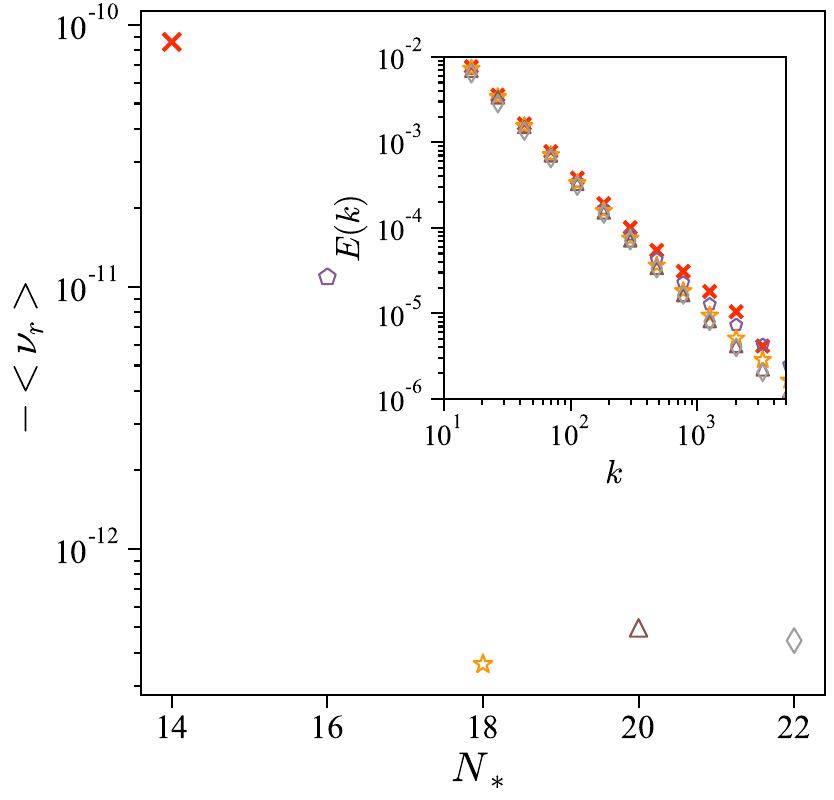}}
                            ~\caption{Convergence of the stochastic protocol for $\mathcal{E} \approx 8.6$. The main panel shows an almost constant $\avg{\nu_r}$ for $N_* \geq 18^3$. The inset presents the corresponding energy spectra highlighting a good collapse for $N_* \geq 20^3$.}
                            \label{fig:Stoch_conv}
            \end{figure}
            \begin{figure*}[!htb]
                    \centering
                    \adjustbox{valign=t}{
                        \begin{minipage}{0.47\textwidth} 
                        \includegraphics[width=\textwidth]{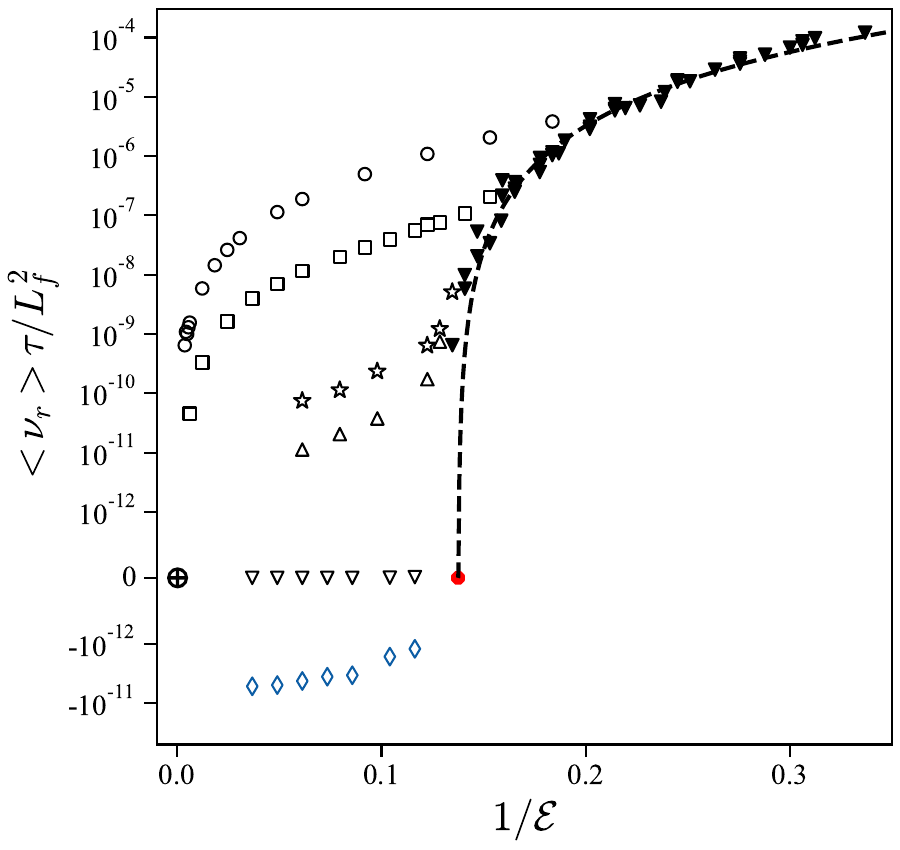}
                        \end{minipage}}
                    \hfill  
                    \adjustbox{valign=t}{\begin{minipage}{0.5\textwidth}
                \includegraphics[width=\textwidth]{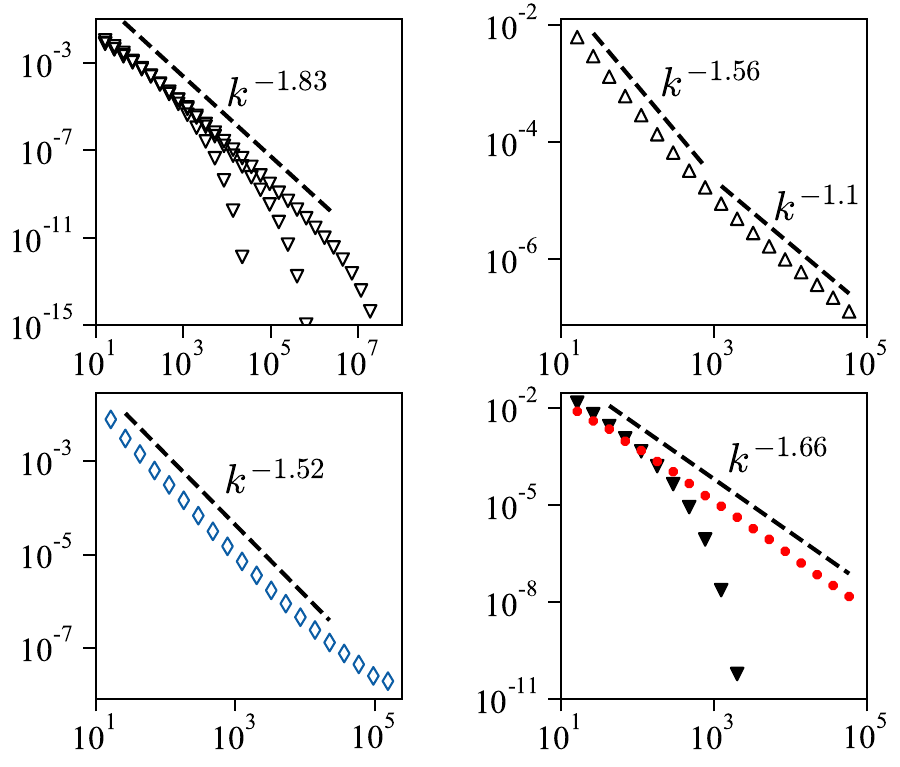}
                \vfill
            \end{minipage}} 
            \put(-125, -214){\resizebox{7pt}{!}{$k$}}
            \put(-266, -105){\begin{turn}{90}
                \resizebox{26pt}{!}{$E(k)$}
                \end{turn}}
            \put(-462, -7){\resizebox{14pt}{!}{\textbf{(a)}}}
            \put(-156, -7){\resizebox{14pt}{!}{\textbf{(b)}}}
            \put(-24, -7){\resizebox{14pt}{!}{\textbf{(c)}}}
            \put(-156, -114){\resizebox{14pt}{!}{\textbf{(d)}}}
            \put(-24, -114){\resizebox{14pt}{!}{\textbf{(e)}}}
        
            \caption{In each figure markers are associated to different grid size $N$ such that \protect\marksymbol{o}{black}{1.5} $N = 8^3$, \protect\marksymbol{square}{black}{1.5} $N = 12^3$, \resizebox{9pt}{!}{\protect$\color{black}\smallstar$}$N = 18^3$, \protect\marksymbol{triangle}{black}{1.8} $N = 20^3$, \protect\diams{black}{1.5} $N = 22^3$, \resizebox{8pt}{!}{\protect\DownTriangle} $N = +\infty$ (adaptative resolution, see Section~\ref{sec:LL}). A different symbol is set for the Euler equations: \resizebox{9pt}{!}{$\protect\oplus$} Euler ($\cal{E} = +\infty$).
             The blue colored symbols correspond to stochastically regularized solutions. \textbf{(a)} Adimensionnalized and averaged reversible viscosity as a function of the inverse of the efficiency $1 / {\cal E}$. The dashed line correspond to a master curve $\avg{\nu_r} \propto (\mathcal{E}^{-1} - \mathcal{E}_*^{-1})^{3}$ , separating the two phases of the system~\cite{costa2023reversible}, where well resolved simulations collapse regardless of the resolution $N$ (full symbols). \textbf{(b)} Snapshots of the energy spectrum taken at three distinct times during the same viscous (RNS) blow-up, illustrating its time evolution. \textbf{(c)} Energy spectrum for truncated RNS - dynamics highlighting thermalization effects. \textbf{(d)} Energy spectrum for a stochastically stabilized blow-up solution. \textbf{(e)} Energy spectra along the master curve of \textbf{(a)}.}
            \label{fig:RNS_blowup}
        \end{figure*}
            \noindent In our simulations, we found convergence, in the sense that $\avg{\nu_r}$ and $E(k)$ no longer depend on $N_*$, for $N_* = 22^3$. An example of this convergence is given in Figure~\ref{fig:Stoch_conv} where the averaged reversible viscosity $\avg{\nu_r}$ is relatively constant and energy spectra indistinguishable $\forall N_* \geq 20^3$.
\section{\label{sec:Results}Results}
            
\begin{figure*}[!htb]
            \centering
            \adjustbox{valign=t}{
                \begin{minipage}{0.48\textwidth} 
                \includegraphics[width=\textwidth]{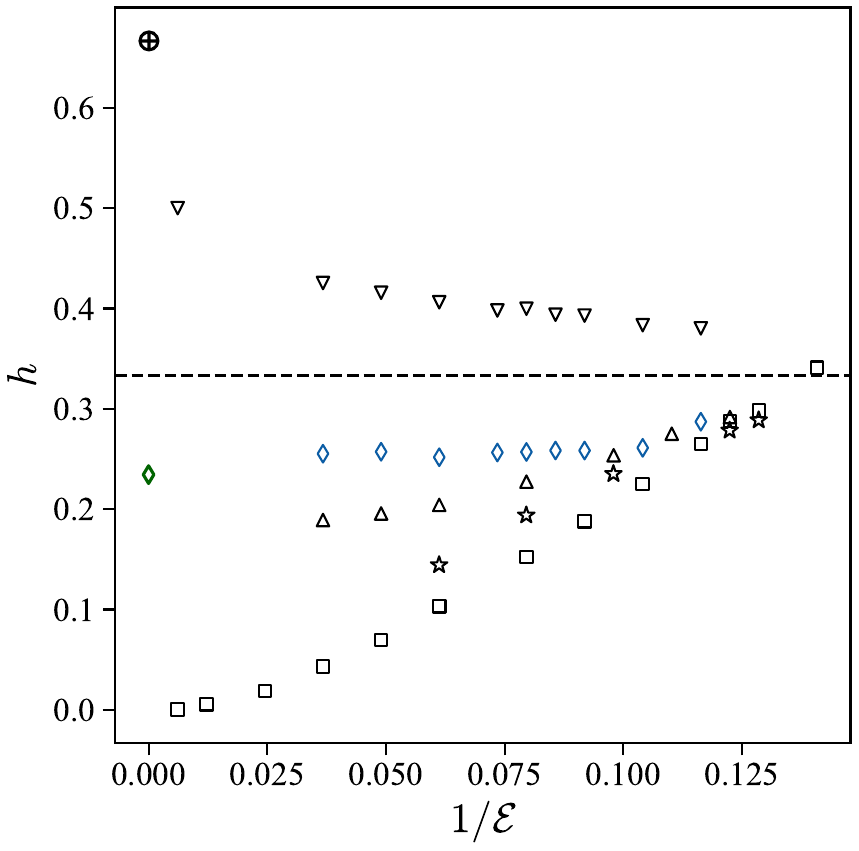}
                \end{minipage}}
            \hfill  
            \adjustbox{valign=t}{\begin{minipage}{0.505\textwidth}
        \includegraphics[width=\textwidth]{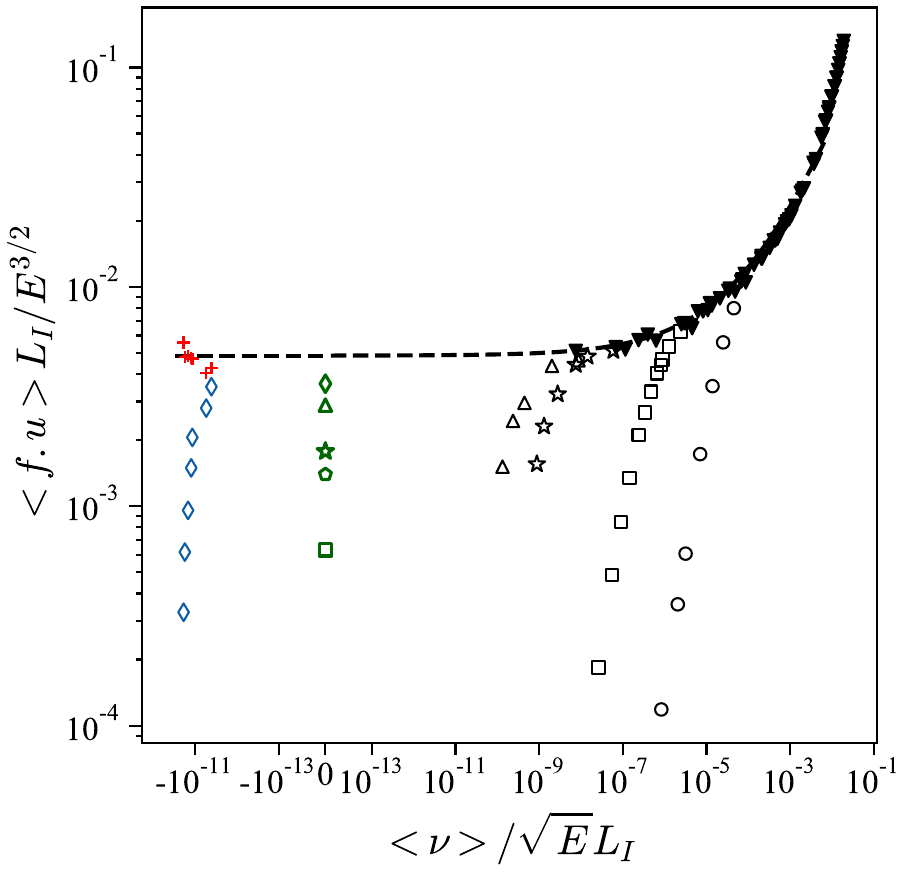}
        \vfill
    \end{minipage}} 
    \put(-510, -0){\resizebox{14pt}{!}{\textbf{(a)}}}
    \put(-257, -0){\resizebox{14pt}{!}{\textbf{(b)}}}
    \put(-365, -15){\resizebox{80pt}{!}{Blow-up solutions}}
    \put(-385, -180){\resizebox{100pt}{!}{Regularized solutions}}

    \caption{Colors and symbols: same as Figure~\ref{fig:RNS_blowup} with the addition of dark green symbols encoding regularized Euler simulations.
    \textbf{(a)} Time averaged spectral exponents as a function of $1 / {\cal E}$. In case of thermalization, only the large scale exponent is reported (e.g extracted from the $-1.56$ slope of Figure~\ref{fig:RNS_blowup}c). The dashed line represent the peculiar value $h_{KG} = 1 / 3$.
    \textbf{(b)} Adimensionnalized mean energy dissipation as a function of the adimensionalized mean viscosity for various solutions of Figure~\ref{fig:RNS_blowup}. For comparison purposes (Euler and RNS) we introduce the integral scale $L_I = \int k^{-1}E(k)dk/\int E(k)dk$. The dashed line represents a fit $D_\epsilon = \epsilon_*\mathcal{E}^{3/2}$, where the value $\cal{E}(<\nu>)$ is obtain by inverting the master curve equation (Figure~\ref{fig:RNS_blowup}a). This fit highlights a possible anomalous dissipation in the inviscid limit which value is compatible with the dissipation found in the regularized Euler simulations (dark green symbols) that slowly converges towards the fit. This value is also compatible with the total energy injection of the regularized RNS solutions (red crosses, \resizebox{7pt}{!}{\textcolor{red}{$\bm{+}$}}) defined by summing the standard energy injections and the viscous contributions (that inject, in average, energy as $\avg{\nu_r}_t < 0$). Small deviations of the red crosses from the dissipative anomaly might be linked to statistical errors as getting really long simulations is difficult, even on LL.}
    \label{fig:Blowup_exp}
    \end{figure*}
    In this article, we consider solutions of the RNSE projected onto a 3D Fourier grid with exponential spacing equal to $\lambda= \phi\approx 1.618$ (the golden number). We use the same code as in~\cite{costa2023reversible} and direct the reader to that paper and the Supplementary Material for further details. Note that this code allows to tackle the case LL-Euler, by simply setting $f_0=\nu_r=0$ at all times. The output of this code provides, for any given efficiency, the time evolution of the reversible viscosity, as well as the Fourier mode amplitudes of the three velocity components, $\hat{u}_i(k,t)$, on the exponential grid. The wavenumber $k_i$ is given by $k_i=k_0\lambda^{n_i}$, where  $i=1,2,3$ and $n_i=1,\dots, N$, with $N$ representing the resolution. Using this data, we compute the energy spectrum $E(k,t)$ at each time step as:
    \begin{eqnarray}
        E(k)&=& \langle\norm{\hat{\bm{u}}}^2\rangle_{S_k},\nonumber\\
        &=&\frac{1}{N_k(\lambda k - k)}\sum_{k\le\norm{\bm{q}}< \lambda k } \norm{\hat{\bm{u}}(\bm{q})}^2,
        \label{eq:defispecm}
    \end{eqnarray}
    where the average $\langle\ \cdot\ \rangle_{S_k}$ is taken over the $N_k\sim (\log k)^{2}$ wave vectors in the shell $S_k$ delimited by spheres of radii $k$ and $\lambda k$. 
    Based on the Isett theorem~\cite{Isett18}, we define a solution of Eqs.~(\ref{eq:RNS}) as non-dissipative if its energy spectrum satisfies $E(k) \propto k^{-1-2h}$ with $h> 1/3$. Hereafter, the parameter $h$ is called the spectral index. Given that solutions may extend from a cutoff of $1/L_f$ to infinity in case of a blow-up, we further constrain the physically admissible solutions to $h>0$, ensuring that the total energy stored in the fluid remains finite. Finally, we observe that for any solution such that $h<1$, the total enstrophy, defined as $\Omega=\int k^2 E(k) dk$ diverges like $k_{max}^{2-2h}$. We therefore call solutions extending to $k_{max}\to\infty$ with $h<1$ "blow up-solutions" or "singular solutions", as their behaviour mimicks that of non-Lifshitz solutions.\ 
    \begin{figure*}[!htb]
        \centering
            \adjustbox{valign=t}{
                \begin{minipage}{0.472\textwidth} 
                \includegraphics[width=\textwidth]{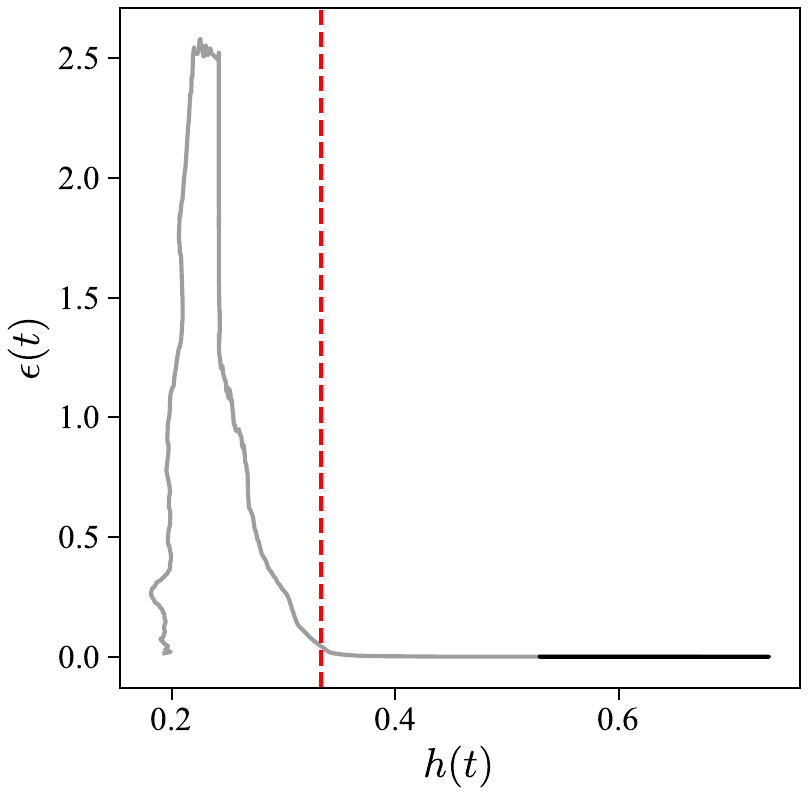}
                \end{minipage}}
            \hfill  
            \adjustbox{valign=t}{\begin{minipage}{0.482\textwidth}
        \includegraphics[width=\textwidth]{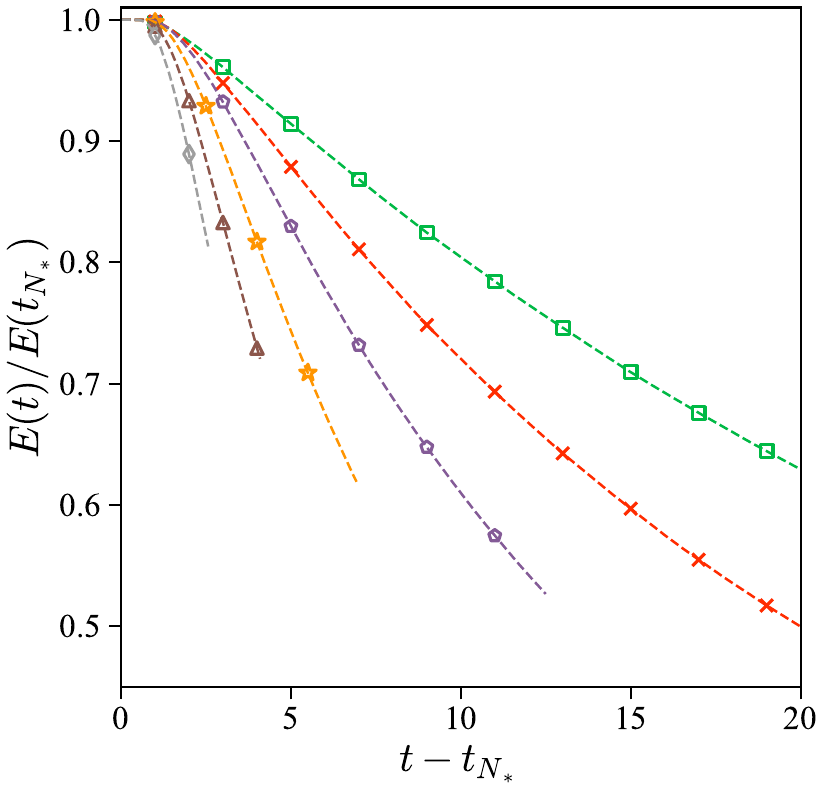}
        \vfill
    \end{minipage}} 
        \put(-506, -0){\resizebox{14pt}{!}{\textbf{(a)}}}
        \put(-247, -0){\resizebox{14pt}{!}{\textbf{(b)}}}
        \caption{Resolutions are encoded by the following symbols and colors \protect\marksymbolfull{square*}{greenplot}{1.5} $N_* = 12^3$, \protect\marksymbol{x}{red}{1.8} $N_* = 14^3$, \protect\marksymbolfull{pentagon*}{purple}{1.5} $N_* = 16^3$, \resizebox{9pt}{!}{\protect$\color{orangeplot}\filledstar$}$N_* = 18^3$, \protect\marksymbolfull{triangle*}{brownplot}{1.8} $N_* = 20^3$, \protect\diamsfull{grey}{1.5} $N_* = 22^3$, \resizebox{9pt}{!}{\color{pinkplot}\protect\RightTriangle} $N_* = 24^3$. \textbf{(a)} Check of the validity of Onsager conjecture on log-lattice: the energy dissipation ($N_* = 22^3$) starts as soon as the spectral exponent goes below $h=1/3$, as materialized by the red vertical dotted line. \textbf{(b)} Time series of the rescaled energy for regularized Euler simulations highlighting an increasing dissipation as $N_* \rightarrow \infty$.}
        \label{fig:DissipReg}
    \end{figure*}

    \subsection{Pre-blow-up solutions} 
        We have performed numerical simulations of Eqs.~(\ref{eq:RNS}) on LL with given forcing $(f_0,L_f)$ inside a triply-periodic cubic domain $\mathcal D$ with side length $2\pi$ using an adaptative grid (see Section~\ref{sec:LL}). The simulations were performed for different values of efficiency, starting with energy distributed across the first few modes.
        The system exhibits two distinct phases, depending on the value of ${\cal E}$:
        (i) A \textit{singular phase} for ${\cal E} > {\cal E}_*$, where ${\cal E}_* \approx 7.3$, associated with finite-time viscous blow-up solutions. These solutions are characterized by a power-law spectrum, $E(k,t)$, which gradually widens towards $k = \infty$ as $\nu_r(t)$ decreases towards $0$ (Figure~\ref{fig:RNS_blowup}b). The asymptotic solution at the finite blow-up time $t_b$ corresponds to a singular solution of the Euler equations, with spectral index depending on the efficiency, as shown in Figure~\ref{fig:Blowup_exp}a. Specifically, the spectral index varies from $2/3$ to $1/3$ as ${\cal E}$ decreases from $\infty$ to ${\cal E}_*$, so that all these solutions are non-dissipative.
        (ii) An \textit{hydrodynamic phase} for ${\cal E} \le {\cal E}_*$, associated with regular, stationary viscous solutions that exhibit a power-law energy spectrum with a spectral index $h \sim 1/3$ at small $k$ and an exponential cut-off at large wavenumbers (Figure~\ref{fig:RNS_blowup}e). Similarly to the singular phase, the viscosity initially decreases steadily, extending the spectrum to larger wavenumbers. However, after a certain time, the viscosity suddenly stops decreasing and instead increases until it reaches a finite value $\nu_*$, at which point the solution becomes statistically stationary. The final value $\nu_*$ depends on ${\cal E}$ and follows the curve $\nu_* \sim (1/{\cal E} - 1/{\cal E}_*)^3$. This curve serves as an attractor for the dynamics when ${\cal E} \le {\cal E}_*$. We verified that these solutions correspond to stationary solutions of the LL-NS equations at fixed viscosity, which are attractive in the following sense: if we freeze the viscosity at some point during the blow-up development (for ${\cal E} > {\cal E}_*$), the solution converges towards the stationary viscous solution of the NSE corresponding to this viscosity by gradually decreasing its efficiency (see Appendix~\ref{app:Stab}). The energy dissipation of these solutions is plotted as a function of viscosity in Figure~\ref{fig:Blowup_exp}b. We observe a steady decrease as $\nu_* \to 0$. Hence, blow-up and post-blow-up solutions at finite efficiency $\mathcal{E}> \mathcal{\mathcal{E}_*}$ can only be studied using the RNS equations.

    \subsection{Post-blow-up solutions via truncation}
        We now focus on blow-up solutions, i.e., for ${\cal E} \ge {\cal E}_*$, which converge to non-dissipative inviscid solutions as $t \to t_b^-$. We will show that these solutions are unstable and transform into different solutions as we go past the blow-up. There are a number of ways to extend the solution beyond the blow-up time. One approach is to freeze the resolution at a fixed number of modes $N$ just before blow-up and let the solution evolve. As the maximum wavenumber $k_{\max}$ is frozen, the entrophy cannot diverge anymore, and the solution is regularized.  For the corresponding solutions,  the viscosity stabilizes at a value $\nu_{th}(N,{\cal E})$ depending on both the resolution and efficiency (Figure~\ref{fig:RNS_blowup}a). The energy cascading from the small wavenumbers accumulates at the largest wavenumbers $k_{\max}$, generating a reflected wave of thermalization that is halted at a finite wavenumber $k_{th}$, which depends on $\nu_{th}$. Below $k_{th}$, the spectrum becomes shallower (Figure~\ref{fig:RNS_blowup}c), resulting in a new exponent $h_{pb}$ (Figure~\ref{fig:Blowup_exp}a, black symbols in the regularized domain). This behavior is the counterpart on LL of what has been observed in direct numerical simulations of the Navier-Stokes equations~\cite{AB20} at much smaller resolutions. Similar dynamics are also observed in the Leith model of turbulence~\cite{connaughton2004warm}, where it corresponds to a shift from an unstable fixed point to a stable one. In our case, the post-blow-up exponent $h_{pb}$ is smaller than the pre-blow-up exponent (Figure~\ref{fig:Blowup_exp}a), and actually becomes less than $1/3$, corresponding to a dissipative solution. For $N = 12^3$, $h_{pb}$ decreases from $1/3$ at ${\cal E} = {\cal E}_*$ to $0$ as ${\cal E} \to \infty$. For higher resolutions, the curve $h_{pb}({\cal E})$ shifts upward and approaches $h_{KG} = 1/3$, corresponding to Kolmogorov dissipative solutions. However, due to computational burden, reaching the infinite-resolution limit of $h_{pb}$ using this regularization approach is difficult. Consequently, we decided to switch to a regularization based on stochastic friction.

    \subsection{Post-blow-up solutions via stochastic friction} 
        The protocol consists in freezing the resolution at a given $N_*$, and introducing a stochastic friction at the largest wavenumbers. Details are provided in Supplementary Material. The corresponding dissipation effectively absorbs all the energy flux from the largest scales, thereby preventing thermalization. After regularization, the dynamics undergo a transient regime, reaching a new statistically stationary state characterized by new values of the mean viscosity $\langle \nu_{pp} \rangle$ and of the local slope $h_{pb}$ (Figure~\ref{fig:RNS_blowup}d). The value of $N_*$ is gradually increased until the results converge. For the efficiency range considered, convergence was achieved at $N_*=22^3$. The corresponding spectral indices $h_{pb}$ are shown in Figure~\ref{fig:Blowup_exp}a (blue colored symbols). These indices are higher than those obtained through thermalization, and it is not clear whether the latter converge to them in the limit $N_*\to\infty$. If they do not, it would suggest that the post-blow-up solution is not unique, consistent with the existence of a stochastic attractor~\cite{mailybaev2024rg, mailybaev2024rgshell}.
        Additionally, the post-blow-up exponents are all lower than the pre-blow-up exponent and increase from $h_{pb}\sim 0.25$ at ${\cal E}=\infty$ (LL-Euler case) to $h_{pb}=1/3$ as ${\cal E}\to {\cal E_*}$. This reinforces the interpretation that the Kolmogorov solution is the attracting fixed point at ${\cal E}= {\cal E_*}$. 

        \medbreak\noindent According to Isett’s theorem, all post-blow-up solutions with spectral exponent  $h_{pb} < \frac{1}{3}$ should exhibit energy dissipation in the limit $N_* \to \infty$. To examine whether this theorem also applies to fluids evolving on LL, we analyze the instantaneous energy variation of a regularized LL-Euler solution ($N_* = 22^3$) as a function of the local spectral exponent $h(t)$, as shown in Figure~\ref{fig:DissipReg}a. We observe that as soon as $h(t)$ crosses the threshold $h = \frac{1}{3}$, the energy starts to decay. This result supports the applicability of Isett's theorem to LL systems and allows us to interpret our findings as follows:
        \begin{itemize}
            \item \textbf{Before blow-up:} The spectral exponents satisfy $h_b > \frac{1}{3}$, corresponding to non-dissipative solutions.
            \item \textbf{After blow-up:} The solutions are characterized by $h_{pb} < \frac{1}{3}$, consistent with the emergence of dissipation through singular dynamics.
        \end{itemize}
        \noindent In the case of LL-Euler ($\mathcal{E} = \infty$), this dissipation is directly evident in the evolution of the total kinetic energy over time (Figure~\ref{fig:DissipReg}b). As $N_* \to \infty$, energy dissipation becomes more pronounced (Figure~\ref{fig:Blowup_exp}b, dark green symbols), even though stochastic friction vanishes in this limit. This behavior indicates that the dissipation originates from the solution itself, rather than from the stochastic protocol.

\begin{figure}[!htb]
            \centering
            \includegraphics[width=0.95\columnwidth]{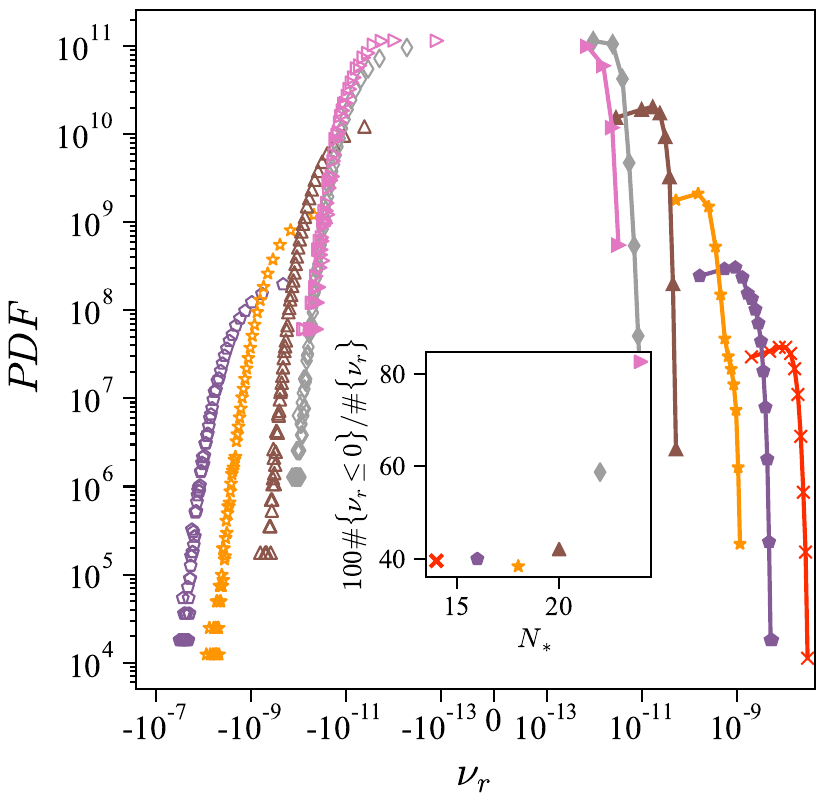}
            \caption{
                Convergence of the stochastic protocol for $\mathcal{E} \approx 8.6$. The main panel shows the average reversible viscosity $\langle \nu_r \rangle$, which becomes nearly constant for $N_* \geq 18^3 $. The inset displays the corresponding energy spectra, demonstrating good collapse for $N_* \geq 20^3 $. The main plot shows the probability distribution of $\nu_r$ for different resolutions. The inset quantifies the percentage of negative viscosity values as a function of $N_*$. Empty symbols represent $\nu_r < 0$, while filled symbols indicate $\nu_r > 0$.
            }
            \label{fig:PDF_nur}
        \end{figure}

        \noindent For finite ${\cal E}$, this effect is evidenced by the behaviour of the viscosity, as $N_*\to \infty$. To ensure energy conservation, the viscosity must assume negative values to compensate for the existence of a dissipative mechanism. Figure~\ref{fig:PDF_nur}, displays the probability distribution of viscosity for ${\cal E}\approx 8.6$ at various resolutions. As the latter increases from $N_* = 8^3$ to $N_* = 24^3$, the proportion of negative values shifts from 50\% to 90\%, with the average positive viscosity being two orders of magnitude smaller than the average negative viscosity. This property holds true whenever ${\cal E}>{\cal E}_*$, as shown in Figure~\ref{fig:RNS_blowup}a (blue diamonds), implying that the mean viscosity after the blow-up always stays negative. 
        To further support the idea that the reversible viscosity compensates for a non-viscous dissipative mechanism, we computed the "total energy injection" for the regularized RNS solutions. This is defined as the sum of the standard energy injection ($\bm{f} \cdot \bm{u}(t)$) and the viscous contribution ($\nu_r(t)\Omega(t)$), with values matching the dissipative anomaly highlighted by the red crosses in Figure~\ref{fig:Blowup_exp}b. Overall, these results support a finite dissipation anomaly, as evidenced by the fit made on Figure~\ref{fig:Blowup_exp}b.
\section{\label{sec:Disc}Discussion}
    We have used fluids on log-lattices and a reversible framework to investigate the influence of the efficiency on the behavior of solutions. We observe a phase transition at a given efficiency, separating regular, viscous solutions, from singular inviscid solutions. The singular solutions experience self-similar blow-up with spectral exponents associated with non-dissipative solutions. Using regularization through truncation or stochastic friction, we go past the blow-up and show that the resulting solutions converge to power-law solutions, with exponents characterizing dissipative solutions. The post-blow up results may depend on the regularization, which would be an indication that they are not unique. We leave this for further work. Globally, the range of spectral exponents we find are comparable to multi-fractal exponents found in ordinary fluids, thereby providing a dynamical scenario for their construction. The occurrence of possible dissipative singular solutions is closely linked to the breaking of symmetries in the equations of motion. In general, the Euler equations are invariant under time translation and time reversal. By Noether theorem, the first symmetry leads to energy conservation. Moreover, time homogeneity means that no special time can be singled out by the dynamics, so that time-reversal can be performed around any time without adding any new information. When a finite-time blow-up occurs, the time translation symmetry is broken, allowing energy to vary. Additionally, applying time-reversal at the blow-up time connects the physical solutions before and after the blow-up. This explains the success of our procedure, based on the reversible Navier-Stokes equations, to unfold the parameter space of the singular Euler equations. By construction, the reversible viscosity changes sign under time-reversal. Therefore, each non-dissipative singular solution of the RNSE with positive viscosity corresponds to a dissipative singular solution of the RNSE with negative viscosity. We have shown how to capture these solutions using truncation or stochastc friction, which allows us to pass the blow-up time. We conjecture that these solutions converge to dissipative singular solutions of the Euler equations. All results have been obtained for fluids projected onto log-lattices. It would be very interesting to test whether these results still hold for fluids on regular Fourier lattices. Indeed, current constructions of dissipative singular solutions usually rely on convex integration methods~\cite{S11}, which are delicate to implement. If our conjecture holds for regular lattices as well, it suggests that the reversible protocol could provide a new way to construct these solutions through a controlled limit procedure. This would likely pave the way for new exact results concerning the existence and properties of dissipative singular solutions of the Euler equations.\
    
\begin{acknowledgments}
This work received funding from the Ecole Polytechnique, from ENS Paris-Saclay and from ANR BANG, grant agreement no. ANR-22-CE30-0025 and ANR TILT grant agreement no. ANR-20-CE30-0035.
\end{acknowledgments}

\appendix
\onecolumngrid

\begin{figure}[!htb]
    \centering
    \begin{minipage}[t]{0.55\textwidth}
        \includegraphics[width=\textwidth]{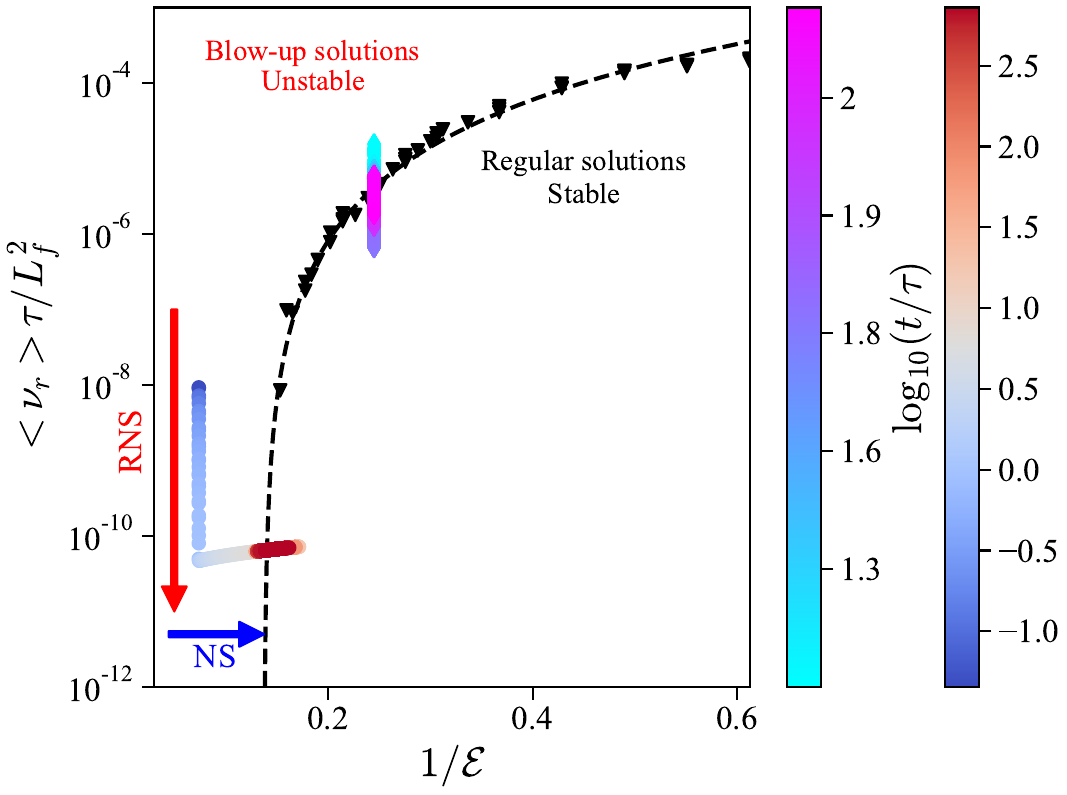}
        \centering
    \end{minipage}
    \hfill
    \begin{minipage}[t]{0.43\textwidth}
        \includegraphics[width=\textwidth]{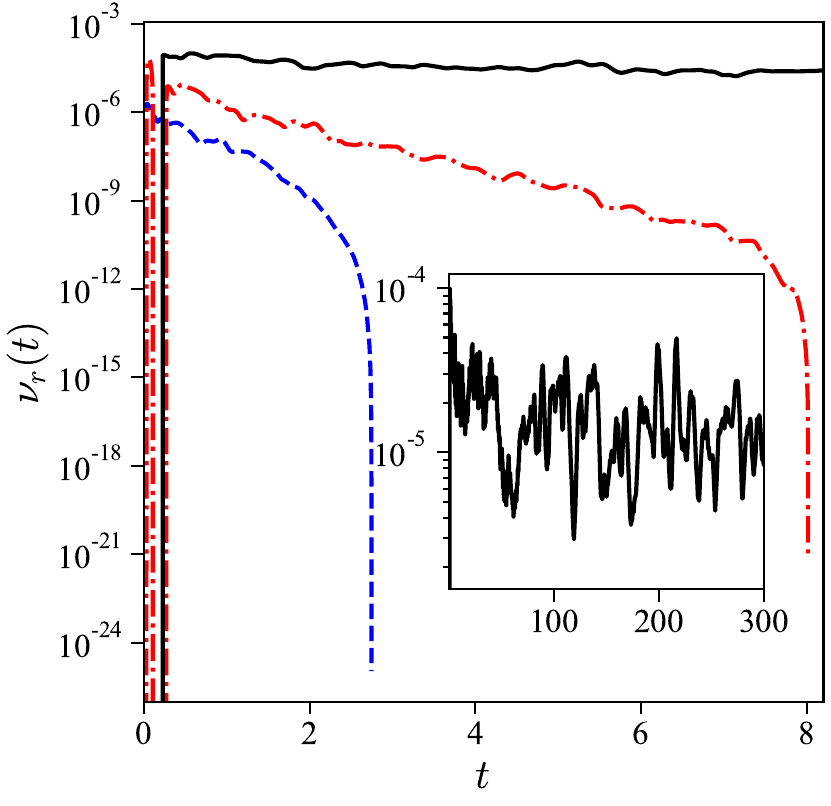}
        \centering
    \end{minipage}
    \put(-510, 205){\resizebox{14pt}{!}{\textbf{(a)}}}
    \put(-220, 205){\resizebox{14pt}{!}{\textbf{(b)}}}
    \caption{\textbf{(a)} Stability of solutions in the two phases. Two trajectories of solutions as a function of time (color-coded) are presented highligthing the instability (resp. stability) of the singular (resp. hydrodynamic) phase. Only a subset of the whole data are plotted here. \textbf{(b)} Time series of the reversible viscosities $\nu_r$ for different efficiencies, illustrating two scenarios: (i) blow-up at high effiency ${\cal E} > {\cal E}^* \approx 7.3$ (\textcolor{red}{\dashdot{ }} $\mathcal{E} \approx 11.7$, \textcolor{blue}{\dashed{ }} $\mathcal{E} \approx 12.6$) (ii) regular solutions for efficiencies lower than ${\cal E}^*$ (\textcolor{black}{\full{ }} $\mathcal{E} \approx 4.1$).
    }
    \label{fig:Stab}
\end{figure}

\twocolumngrid
    
\section{\label{app:Stab}Stability of solutions}
    Both the Euler and RNS equations exhibit blow-up solutions in the range ${\cal E} > {\cal E}^*$. In order to go beyond the blow-up time $t_b$, one might be tempted to simply fix the resolution $N$.
    However, it has already been shown that under resolved solutions undergo thermalization at small scales~\cite{krstulovic2008two}. Through this thermalization, solutions reach a stable "thermalized" branch, which corresponds to truncation effects and thus strongly depends on $N$.
    \medbreak\noindent Once reached, this thermalized branch is stable by switching between NSE and RNSE as a consequence of the Gallavotti conjecture~\cite{gallavotti1996equivalence, costa2023reversible}. However, the instanteneous RNS solutions (i.e non-thermalized) are unstable under such switch, see Figure~\ref{fig:Stab}a. The solution then converges to the stable branch $(1/{\cal E} - 1/{\cal E^*})^3$. 
    \medbreak\noindent In addition, performing RNS simulations with adaptative resolution in the hydrodynamic phase yields the same solutions as the one obtained with fixed, sufficiently large, resolution. Figure~\ref{fig:Stab}b presents the time series of the reversible viscosities highlitighting blow-ups in the unstable phase (blue and green lines) as $\nu_r \propto \Omega^{-1}$. The orange curve shows that there are no blow-ups in the hydrodynamic phase as $\nu_r$ achieves a steady state. Note that the amplitude of the oscillations are correlated to the position in the phase transition that the system exhibits~\cite{shukla2019phase, costa2023reversible}.

\section{Spectral indices extraction}
\begin{figure} [!htb]   
        {\includegraphics[width =.78\columnwidth]{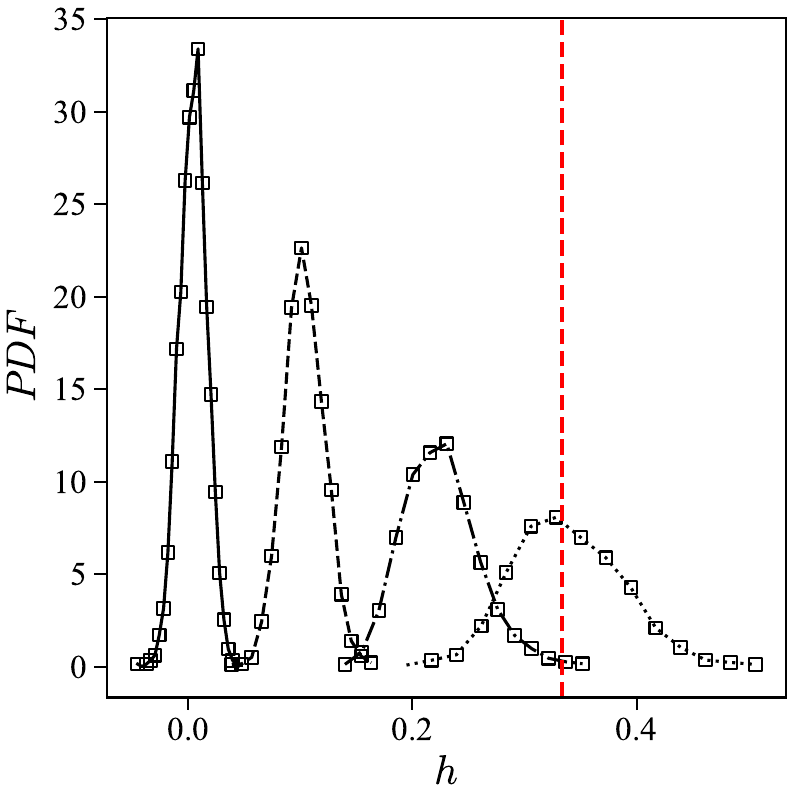}}
            ~\caption{PDFs of the spectral index $h$ for various efficiencies $\cal E$,  \textcolor{black}{\full{ }} $\mathcal{E} \approx 4e3 $, \textcolor{black}{\dashed{ }} $\mathcal{E} \approx 83.3 $, \textcolor{black}{\dashdot{ }} $\mathcal{E} \approx 32.3 $, \textcolor{black}{\dotted{ }} $\mathcal{E} \approx 0.86 $. The marker \protect\marksymbol{square}{black}{1.5} refers to the resolution, here $N = 12^3$. The vertical red dashed line correspond to the Kolmogorov value $h_{KG} = 1/3$.}
        \label{fig:h_pdf}
    \end{figure}

    From the energy spectrum $E(k,t)$, we define the \textit{spectral index} $h$ such that $E(k,t) \propto k^{-1-2h}$. Spectral indices can thus be directly extracted from the slope of the energy spectrum in the inertial range (see Figure~1b-e of the main text for examples of spectra). Note that in presence of truncation effects, the energy spectrum exhibits two slopes: (i) in the inertial range, associated with the spectral index $h$ and (ii) at high $k$ associated with the Galerkin truncation (see Figure~1c of the main text). Examples of PDF of the spectral index $h$, for various efficiencies, are given in Figure~\ref{fig:h_pdf}.
\clearpage
\providecommand{\noopsort}[1]{}\providecommand{\singleletter}[1]{#1}%


\begin{thebibliography}{20}%
\makeatletter
\providecommand \@ifxundefined [1]{%
 \@ifx{#1\undefined}
}%
\providecommand \@ifnum [1]{%
 \ifnum #1\expandafter \@firstoftwo
 \else \expandafter \@secondoftwo
 \fi
}%
\providecommand \@ifx [1]{%
 \ifx #1\expandafter \@firstoftwo
 \else \expandafter \@secondoftwo
 \fi
}%
\providecommand \natexlab [1]{#1}%
\providecommand \enquote  [1]{``#1''}%
\providecommand \bibnamefont  [1]{#1}%
\providecommand \bibfnamefont [1]{#1}%
\providecommand \citenamefont [1]{#1}%
\providecommand \href@noop [0]{\@secondoftwo}%
\providecommand \href [0]{\begingroup \@sanitize@url \@href}%
\providecommand \@href[1]{\@@startlink{#1}\@@href}%
\providecommand \@@href[1]{\endgroup#1\@@endlink}%
\providecommand \@sanitize@url [0]{\catcode `\\12\catcode `\$12\catcode `\&12\catcode `\#12\catcode `\^12\catcode `\_12\catcode `\%12\relax}%
\providecommand \@@startlink[1]{}%
\providecommand \@@endlink[0]{}%
\providecommand \url  [0]{\begingroup\@sanitize@url \@url }%
\providecommand \@url [1]{\endgroup\@href {#1}{\urlprefix }}%
\providecommand \urlprefix  [0]{URL }%
\providecommand \Eprint [0]{\href }%
\providecommand \doibase [0]{https://doi.org/}%
\providecommand \selectlanguage [0]{\@gobble}%
\providecommand \bibinfo  [0]{\@secondoftwo}%
\providecommand \bibfield  [0]{\@secondoftwo}%
\providecommand \translation [1]{[#1]}%
\providecommand \BibitemOpen [0]{}%
\providecommand \bibitemStop [0]{}%
\providecommand \bibitemNoStop [0]{.\EOS\space}%
\providecommand \EOS [0]{\spacefactor3000\relax}%
\providecommand \BibitemShut  [1]{\csname bibitem#1\endcsname}%
\let\auto@bib@innerbib\@empty
\bibitem [{\citenamefont {Onsager}(1949)}]{onsager1949statistical}%
  \BibitemOpen
  \bibfield  {author} {\bibinfo {author} {\bibfnamefont {L.}~\bibnamefont {Onsager}},\ }\bibfield  {title} {\bibinfo {title} {Statistical hydrodynamics},\ }\href@noop {} {\bibfield  {journal} {\bibinfo  {journal} {Il Nuovo Cimento (1943-1954)}\ }\textbf {\bibinfo {volume} {6}},\ \bibinfo {pages} {279} (\bibinfo {year} {1949})}\BibitemShut {NoStop}%
\bibitem [{\citenamefont {Buckmaster}\ \emph {et~al.}(2018)\citenamefont {Buckmaster}, \citenamefont {De~Lellis}, \citenamefont {Székelyhidi},\ and\ \citenamefont {Vicol}}]{Buckmaster18}%
  \BibitemOpen
  \bibfield  {author} {\bibinfo {author} {\bibfnamefont {T.}~\bibnamefont {Buckmaster}}, \bibinfo {author} {\bibfnamefont {C.}~\bibnamefont {De~Lellis}}, \bibinfo {author} {\bibfnamefont {L.}~\bibnamefont {Székelyhidi}},\ and\ \bibinfo {author} {\bibfnamefont {V.}~\bibnamefont {Vicol}},\ }\bibfield  {title} {\bibinfo {title} {Onsager’s conjecture for admissible weak solutions},\ }\href {https://onlinelibrary.wiley.com/doi/abs/10.1002/cpa.21781} {\bibfield  {journal} {\bibinfo  {journal} {Communications on Pure and Applied Mathematics}\ }\textbf {\bibinfo {volume} {0}} (\bibinfo {year} {2018})},\ \Eprint {https://arxiv.org/abs/https://onlinelibrary.wiley.com/doi/pdf/10.1002/cpa.21781} {https://onlinelibrary.wiley.com/doi/pdf/10.1002/cpa.21781} \BibitemShut {NoStop}%
\bibitem [{\citenamefont {Arneodo}\ \emph {et~al.}(1996)\citenamefont {Arneodo}, \citenamefont {Baudet}, \citenamefont {Belin}, \citenamefont {Benzi}, \citenamefont {Castaing}, \citenamefont {Chabaud}, \citenamefont {Chavarria}, \citenamefont {Ciliberto}, \citenamefont {Camussi}, \citenamefont {Chillà} \emph {et~al.}}]{A96}%
  \BibitemOpen
  \bibfield  {author} {\bibinfo {author} {\bibfnamefont {A.}~\bibnamefont {Arneodo}}, \bibinfo {author} {\bibfnamefont {C.}~\bibnamefont {Baudet}}, \bibinfo {author} {\bibfnamefont {F.}~\bibnamefont {Belin}}, \bibinfo {author} {\bibfnamefont {R.}~\bibnamefont {Benzi}}, \bibinfo {author} {\bibfnamefont {B.}~\bibnamefont {Castaing}}, \bibinfo {author} {\bibfnamefont {B.}~\bibnamefont {Chabaud}}, \bibinfo {author} {\bibfnamefont {R.}~\bibnamefont {Chavarria}}, \bibinfo {author} {\bibfnamefont {S.}~\bibnamefont {Ciliberto}}, \bibinfo {author} {\bibfnamefont {R.}~\bibnamefont {Camussi}}, \bibinfo {author} {\bibfnamefont {F.}~\bibnamefont {Chillà}}, \emph {et~al.},\ }\bibfield  {title} {\bibinfo {title} {Structure functions in turbulence, in various flow configurations, at reynolds number between 30 and 5000, using extended self-similarity},\ }\href {https://doi.org/10.1209/epl/i1996-00472-2} {\bibfield  {journal} {\bibinfo  {journal} {Europhys. Lett.}\ }\textbf {\bibinfo {volume} {34}},\ \bibinfo {pages} {411}
  (\bibinfo {year} {1996})}\BibitemShut {NoStop}%
\bibitem [{\citenamefont {Faller}(2022)}]{faller2022dissipation}%
  \BibitemOpen
  \bibfield  {author} {\bibinfo {author} {\bibfnamefont {H.}~\bibnamefont {Faller}},\ }\emph {\bibinfo {title} {Dissipation in Turbulent Flows}},\ \href@noop {} {Ph.D. thesis},\ \bibinfo  {school} {Université Paris-Saclay} (\bibinfo {year} {2022})\BibitemShut {NoStop}%
\bibitem [{\citenamefont {Isett}(2018)}]{Isett18}%
  \BibitemOpen
  \bibfield  {author} {\bibinfo {author} {\bibfnamefont {P.}~\bibnamefont {Isett}},\ }\bibfield  {title} {\bibinfo {title} {A proof of onsager's conjecture},\ }\href {https://doi.org/10.4007/annals.2018.188.3.4} {\bibfield  {journal} {\bibinfo  {journal} {Annals of Mathematics}\ }\textbf {\bibinfo {volume} {188}},\ \bibinfo {pages} {871} (\bibinfo {year} {2018})}\BibitemShut {NoStop}%
\bibitem [{\citenamefont {Campolina}(2019)}]{campolina2019fluid}%
  \BibitemOpen
  \bibfield  {author} {\bibinfo {author} {\bibfnamefont {C.}~\bibnamefont {Campolina}},\ }\emph {\bibinfo {title} {Fluid models for weak turbulence and energy cascades}},\ \href@noop {} {Master's thesis},\ \bibinfo  {school} {IMPA} (\bibinfo {year} {2019})\BibitemShut {NoStop}%
\bibitem [{\citenamefont {Campolina}\ and\ \citenamefont {Mailybaev}(2021)}]{campolina2021fluid}%
  \BibitemOpen
  \bibfield  {author} {\bibinfo {author} {\bibfnamefont {C.~S.}\ \bibnamefont {Campolina}}\ and\ \bibinfo {author} {\bibfnamefont {A.~A.}\ \bibnamefont {Mailybaev}},\ }\bibfield  {title} {\bibinfo {title} {Fluid model for the self-similar evolution of weakly turbulent solutions of shell models},\ }\href@noop {} {\bibfield  {journal} {\bibinfo  {journal} {Nonlinearity}\ }\textbf {\bibinfo {volume} {34}},\ \bibinfo {pages} {4684} (\bibinfo {year} {2021})}\BibitemShut {NoStop}%
\bibitem [{\citenamefont {Campolina}\ and\ \citenamefont {Mailybaev}(2018)}]{campolina2018chaotic}%
  \BibitemOpen
  \bibfield  {author} {\bibinfo {author} {\bibfnamefont {C.~S.}\ \bibnamefont {Campolina}}\ and\ \bibinfo {author} {\bibfnamefont {A.~A.}\ \bibnamefont {Mailybaev}},\ }\bibfield  {title} {\bibinfo {title} {Chaotic blowup in a shell model of convective turbulence},\ }\href@noop {} {\bibfield  {journal} {\bibinfo  {journal} {Physical Review Letters}\ }\textbf {\bibinfo {volume} {121}},\ \bibinfo {pages} {064501} (\bibinfo {year} {2018})}\BibitemShut {NoStop}%
\bibitem [{\citenamefont {Pikeroen}\ \emph {et~al.}(2024)\citenamefont {Pikeroen}, \citenamefont {Barral}, \citenamefont {Costa}, \citenamefont {Campolina}, \citenamefont {Mailybaev},\ and\ \citenamefont {Dubrulle}}]{pikeroenSingularite}%
  \BibitemOpen
  \bibfield  {author} {\bibinfo {author} {\bibfnamefont {Q.}~\bibnamefont {Pikeroen}}, \bibinfo {author} {\bibfnamefont {A.}~\bibnamefont {Barral}}, \bibinfo {author} {\bibfnamefont {G.}~\bibnamefont {Costa}}, \bibinfo {author} {\bibfnamefont {C.}~\bibnamefont {Campolina}}, \bibinfo {author} {\bibfnamefont {A.~A.}\ \bibnamefont {Mailybaev}},\ and\ \bibinfo {author} {\bibfnamefont {B.}~\bibnamefont {Dubrulle}},\ }\bibfield  {title} {\bibinfo {title} {Singularité en turbulence réversible: une approche renormalisation-invariante},\ }\href@noop {} {\bibfield  {journal} {\bibinfo  {journal} {Nonlinearity}\ }\textbf {\bibinfo {volume} {37}},\ \bibinfo {pages} {115003} (\bibinfo {year} {2024})}\BibitemShut {NoStop}%
\bibitem [{\citenamefont {Gallavotti}(1996)}]{gallavotti1996equivalence}%
  \BibitemOpen
  \bibfield  {author} {\bibinfo {author} {\bibfnamefont {G.}~\bibnamefont {Gallavotti}},\ }\bibfield  {title} {\bibinfo {title} {Equivalence of dynamical ensembles and navier-stokes equations},\ }\href@noop {} {\bibfield  {journal} {\bibinfo  {journal} {Physics Letters A}\ }\textbf {\bibinfo {volume} {223}},\ \bibinfo {pages} {91} (\bibinfo {year} {1996})}\BibitemShut {NoStop}%
\bibitem [{\citenamefont {Shukla}\ \emph {et~al.}(2019)\citenamefont {Shukla}, \citenamefont {Dubrulle}, \citenamefont {Nazarenko}, \citenamefont {Krstulovic},\ and\ \citenamefont {Thalabard}}]{shukla2019phase}%
  \BibitemOpen
  \bibfield  {author} {\bibinfo {author} {\bibfnamefont {V.}~\bibnamefont {Shukla}}, \bibinfo {author} {\bibfnamefont {B.}~\bibnamefont {Dubrulle}}, \bibinfo {author} {\bibfnamefont {S.}~\bibnamefont {Nazarenko}}, \bibinfo {author} {\bibfnamefont {G.}~\bibnamefont {Krstulovic}},\ and\ \bibinfo {author} {\bibfnamefont {S.}~\bibnamefont {Thalabard}},\ }\bibfield  {title} {\bibinfo {title} {Phase transitions in non-equilibrium turbulence},\ }\href@noop {} {\bibfield  {journal} {\bibinfo  {journal} {Physical Review E}\ }\textbf {\bibinfo {volume} {100}},\ \bibinfo {pages} {043104} (\bibinfo {year} {2019})}\BibitemShut {NoStop}%
\bibitem [{\citenamefont {Costa}\ \emph {et~al.}(2023)\citenamefont {Costa}, \citenamefont {Barral},\ and\ \citenamefont {Dubrulle}}]{costa2023reversible}%
  \BibitemOpen
  \bibfield  {author} {\bibinfo {author} {\bibfnamefont {G.}~\bibnamefont {Costa}}, \bibinfo {author} {\bibfnamefont {A.}~\bibnamefont {Barral}},\ and\ \bibinfo {author} {\bibfnamefont {B.}~\bibnamefont {Dubrulle}},\ }\bibfield  {title} {\bibinfo {title} {Reversible turbulence and its statistical mechanics},\ }\href@noop {} {\bibfield  {journal} {\bibinfo  {journal} {Physical Review E}\ }\textbf {\bibinfo {volume} {107}},\ \bibinfo {pages} {065106} (\bibinfo {year} {2023})}\BibitemShut {NoStop}%
\bibitem [{\citenamefont {Lopez}\ \emph {et~al.}(2025)\citenamefont {Lopez}, \citenamefont {Costa}, \citenamefont {Barral}, \citenamefont {Pikeroen}, \citenamefont {Shukla},\ and\ \citenamefont {Dubrulle}}]{Lopez2025}%
  \BibitemOpen
  \bibfield  {author} {\bibinfo {author} {\bibfnamefont {A.}~\bibnamefont {Lopez}}, \bibinfo {author} {\bibfnamefont {G.}~\bibnamefont {Costa}}, \bibinfo {author} {\bibfnamefont {A.}~\bibnamefont {Barral}}, \bibinfo {author} {\bibfnamefont {Q.}~\bibnamefont {Pikeroen}}, \bibinfo {author} {\bibfnamefont {V.}~\bibnamefont {Shukla}},\ and\ \bibinfo {author} {\bibfnamefont {B.}~\bibnamefont {Dubrulle}},\ }\bibfield  {title} {\bibinfo {title} {Efficiency of turbulence},\ }\href {https://arxiv.org/abs/2508.05686} {\bibfield  {journal} {\bibinfo  {journal} {arXiv preprint arXiv:2508.05686}\ } (\bibinfo {year} {2025})}\BibitemShut {NoStop}%
\bibitem [{\citenamefont {Krstulovic}\ and\ \citenamefont {Brachet}(2008)}]{krstulovic2008two}%
  \BibitemOpen
  \bibfield  {author} {\bibinfo {author} {\bibfnamefont {G.}~\bibnamefont {Krstulovic}}\ and\ \bibinfo {author} {\bibfnamefont {M.-E.}\ \bibnamefont {Brachet}},\ }\bibfield  {title} {\bibinfo {title} {Two-fluid dynamics of the gross–pitaevskii condensate in the classical-field approximation},\ }\href@noop {} {\bibfield  {journal} {\bibinfo  {journal} {Physica D: Nonlinear Phenomena}\ }\textbf {\bibinfo {volume} {237}},\ \bibinfo {pages} {2015} (\bibinfo {year} {2008})}\BibitemShut {NoStop}%
\bibitem [{\citenamefont {Murugan}\ \emph {et~al.}(2020)\citenamefont {Murugan}, \citenamefont {Frisch}, \citenamefont {Nazarenko}, \citenamefont {Besse},\ and\ \citenamefont {Ray}}]{murugan2020suppressing}%
  \BibitemOpen
  \bibfield  {author} {\bibinfo {author} {\bibfnamefont {S.~D.}\ \bibnamefont {Murugan}}, \bibinfo {author} {\bibfnamefont {U.}~\bibnamefont {Frisch}}, \bibinfo {author} {\bibfnamefont {S.}~\bibnamefont {Nazarenko}}, \bibinfo {author} {\bibfnamefont {N.}~\bibnamefont {Besse}},\ and\ \bibinfo {author} {\bibfnamefont {S.~S.}\ \bibnamefont {Ray}},\ }\bibfield  {title} {\bibinfo {title} {Suppressing blow-up in shell models of turbulence by spectral truncation},\ }\href@noop {} {\bibfield  {journal} {\bibinfo  {journal} {Physical Review Research}\ }\textbf {\bibinfo {volume} {2}},\ \bibinfo {pages} {033202} (\bibinfo {year} {2020})}\BibitemShut {NoStop}%
\bibitem [{\citenamefont {Alexakis}\ and\ \citenamefont {Brachet}(2020)}]{AB20}%
  \BibitemOpen
  \bibfield  {author} {\bibinfo {author} {\bibfnamefont {A.}~\bibnamefont {Alexakis}}\ and\ \bibinfo {author} {\bibfnamefont {M.-E.}\ \bibnamefont {Brachet}},\ }\bibfield  {title} {\bibinfo {title} {Impact of non-local interactions on turbulent cascades},\ }\href@noop {} {\bibfield  {journal} {\bibinfo  {journal} {Journal of Fluid Mechanics}\ }\textbf {\bibinfo {volume} {884}},\ \bibinfo {pages} {A33} (\bibinfo {year} {2020})}\BibitemShut {NoStop}%
\bibitem [{\citenamefont {Connaughton}\ and\ \citenamefont {Nazarenko}(2004)}]{connaughton2004warm}%
  \BibitemOpen
  \bibfield  {author} {\bibinfo {author} {\bibfnamefont {C.}~\bibnamefont {Connaughton}}\ and\ \bibinfo {author} {\bibfnamefont {S.}~\bibnamefont {Nazarenko}},\ }\bibfield  {title} {\bibinfo {title} {Warm cascades and anomalous scaling in a diffusion model of turbulence},\ }\href@noop {} {\bibfield  {journal} {\bibinfo  {journal} {Physical Review Letters}\ }\textbf {\bibinfo {volume} {92}},\ \bibinfo {pages} {044501} (\bibinfo {year} {2004})}\BibitemShut {NoStop}%
\bibitem [{\citenamefont {Mailybaev}(2024{\natexlab{a}})}]{mailybaev2024rg}%
  \BibitemOpen
  \bibfield  {author} {\bibinfo {author} {\bibfnamefont {A.~A.}\ \bibnamefont {Mailybaev}},\ }\href@noop {} {\bibinfo {title} {Renormalization group formulation for turbulence}} (\bibinfo {year} {2024}{\natexlab{a}}),\ \bibinfo {note} {arXiv preprint arXiv:2410.14903},\ \Eprint {https://arxiv.org/abs/2410.14903} {arXiv:2410.14903} \BibitemShut {NoStop}%
\bibitem [{\citenamefont {Mailybaev}(2024{\natexlab{b}})}]{mailybaev2024rgshell}%
  \BibitemOpen
  \bibfield  {author} {\bibinfo {author} {\bibfnamefont {A.~A.}\ \bibnamefont {Mailybaev}},\ }\href@noop {} {\bibinfo {title} {Renormalization group analysis of shell models of turbulence}} (\bibinfo {year} {2024}{\natexlab{b}}),\ \bibinfo {note} {arXiv preprint arXiv:2408.04659},\ \Eprint {https://arxiv.org/abs/2408.04659} {arXiv:2408.04659} \BibitemShut {NoStop}%
\bibitem [{\citenamefont {Székelyhidi}(2011)}]{S11}%
  \BibitemOpen
  \bibfield  {author} {\bibinfo {author} {\bibfnamefont {L.}~\bibnamefont {Székelyhidi}},\ }\bibfield  {title} {\bibinfo {title} {Weak solutions to the incompressible euler equations with vortex sheet initial data},\ }\href {https://www.sciencedirect.com/science/article/pii/S1631073X11002639} {\bibfield  {journal} {\bibinfo  {journal} {Comptes Rendus Mathematique}\ }\textbf {\bibinfo {volume} {349}},\ \bibinfo {pages} {1063} (\bibinfo {year} {2011})}\BibitemShut {NoStop}%
\end{thebibliography}
\end{document}